\documentclass[12pt, a4paper]{article}
\pdfoutput=1

\usepackage{amsmath}
\usepackage{amsfonts}
\usepackage{amssymb}
\usepackage{graphicx, rotating}
\usepackage{epstopdf}
\usepackage{epsfig}
\usepackage{latexsym}
\usepackage{graphicx}
\usepackage{color}
\usepackage{amsmath,amssymb}
\usepackage{cite}
\usepackage{slashed}
\usepackage{hyperref}

\numberwithin{equation}{section}

\hypersetup{colorlinks, citecolor=bluscuro, linkcolor=bluscuro, urlcolor=bluscuro}
\definecolor{rossos}{rgb}{0.8,0.2,0.3}
\definecolor{bluscuro}{rgb}{0.15, 0.2, 0.9}

\makeatother   % Cancel the effect of \makeatletter
\newcommand{\GeV}{{\rm \,GeV}}
\newcommand{\TeV}{{\rm \,TeV}}

\def\tm{\widetilde m}
\def\tbe{\tan\beta}

 \def\be   {\begin{equation}}   \def\ee   {\end{equation}}
 \def\ba   {\begin{array}}      \def\ea   {\end{array}}
 \def\bea  {\begin{eqnarray}}   \def\eea  {\end{eqnarray}}
 \def\bean {\begin{eqnarray*}}  \def\eean {\end{eqnarray*}}
 \def\nn{\nonumber}

\begin{document}

\begin{flushright}
 \hfill
 {\footnotesize CERN-PH-TH/2011-177  \\[0.1cm] UAB-FT-695\\ ULB-TH/11-18}
\end{flushright}

\vspace{0.5cm}
\begin{center}

\def\thefootnote{\fnsymbol{footnote}}

{\Large \bf Magnetic Fields at First Order Phase Transition:\\
  \vskip 0.3cm A Threat to Electroweak Baryogenesis}
\\[1.2cm]
{\large \textsc{Andrea De Simone} $^{\rm a}$, \textsc{Germano Nardini} $^{\rm b}$,} \\[0.2cm]
{\large \textsc{Mariano Quir\'os} $^{\rm c}$, \textsc{Antonio
    Riotto} $^{\rm d,e}$}
\\[1cm]

{\small \textit{$^{\rm a}$ Institut de Th\'eorie des Ph\'enom\`enes Physiques,\\
 \'Ecole Polytechnique F\'ed\'erale de Lausanne, CH-1015 Lausanne,
Switzerland}}

{\small \textit{$^{\rm b}$ Service de Physique Th\'eorique,
    Universit\'e Libre de Bruxelles, 1050 Brussels, Belgium}}

{\small \textit{$^{\rm c}$ ICREA, Instituci\`o Catalana de Recerca i
    Estudis Avan\c{c}ats, and\\IFAE, Universitat Aut\`onoma de Barcelona,
    08193 Bellaterra, Barcelona, Spain}}

{\small \textit{$^{\rm d}$  CERN, PH-TH Division, CH-1211,
Gen\`eve 23,  Switzerland}}

{\small \textit{$^{\rm e}$  INFN, Sezione di Padova, Via Marzolo 8, I-35131, Padova, Italy}}

\vspace{.2cm}

\end{center}

\vspace{.8cm}

\begin{center}
\textbf{Abstract}
\end{center}

\noindent
The generation of the observed baryon asymmetry may have taken place
during the electroweak phase transition, thus involving physics
testable at LHC, a scenario dubbed electroweak baryogenesis.  In this
paper we point out that the magnetic field which is produced in the
bubbles of a first order phase transition endangers the baryon
asymmetry produced in the bubble walls. The reason being that the
produced magnetic field couples to the sphaleron magnetic moment
and lowers the sphaleron energy; this strengthens the sphaleron
transitions inside the bubbles and triggers a more effective wash out
of the baryon asymmetry. We apply this scenario to the Minimal
Supersymmetric extension of the Standard Model (MSSM) where, in the absence of a magnetic field,  successful
electroweak baryogenesis requires the lightest CP-even Higgs and the
right-handed stop masses to be lighter than about 127 GeV and 120 GeV,
respectively. We show that even for moderate values of the magnetic
field, the Higgs mass required to preserve the baryon asymmetry is
below the present experimental bound.  As a consequence electroweak baryogenesis within
the MSSM should be confronted on the one hand to future measurements
at the LHC on the Higgs and the right-handed stop masses, and on the
other hand to more precise calculations of the magnetic field produced
at the electroweak phase transition.

\def\thefootnote{\arabic{footnote}}
\setcounter{footnote}{0}
\pagestyle{empty}

\newpage
\pagestyle{plain}
\setcounter{page}{1}

%%%%%%%%%%%%%%%%%%%%%%%%%%%%%%%%%%%%%
\section{Introduction}
%%%%%%%%%%%%%%%%%%%%%%%%%%%%%%%%%%%%%
\noindent
Electroweak baryogenesis (EWBG) ~\cite{baryogenesis,reviews} is a very
elegant mechanism for generating the baryon asymmetry of the Universe
(BAU). It relies on physics at the weak scale and can therefore be
tested at present accelerator energies, in particular at the
LHC. During the electroweak phase transition (EWPT) bubbles of the
broken phase are nucleated and expand. Particles in the plasma are
reflected off the bubble walls where CP is violated and CP-violating
currents may be generated. If the currents efficiently diffuse into
the unbroken phase they may be converted into a baryon asymmetry by
the action of the baryon number violating sphaleron
processes~\cite{sphalerons}. The baryon asymmetry then flows into the
interior of the bubble where it is preserved provided that the
sphaleron interactions are sufficiently switched off in the broken
phase which defines a strong enough first order phase transition.

In this work we point out an effect that seems to have passed
unnoticed so far and that may require a stronger 
first order phase transition, for EWBG purposes. 
During the first order phase transition magnetic fields are
unavoidably generated~\cite{ereview}.  Bubble collisions generate a
level of turbulence and hence vorticity in the fluid. The turbulent
conducting fluid develops magnetic turbulence resulting in magnetic
fields on all scale sizes. The turbulence in the fluid amplifies
whatever seed fields are present to finite-amplitude large-scale size
magnetic fields \cite{Stevens}.  The relevant time scale for the amplification of
fields on length scale $\ell$ is of order $(\ell/R_{\rm b})t_{\rm
  pt}$, where $R_{\rm b}$ is the radius of the bubble moving with
velocity $v_{\rm w}$ and $t_{\rm pt}\sim R_{\rm b}/v_{\rm w}$ is the
duration of the phase transition. If the field growth is exponential 
fields on scales $\ell\lesssim R_{\rm b}$
can be amplified by many $e$-folds~\cite{baym,sigl}. When the magnetic
turbulence becomes fully developed the kinetic energy of the
turbulent flow is equipartitioned with that of the magnetic field
energy implying that the magnetic fields $B(R_{\rm b})$ on the size
of the bubble radius is $B^2/2={\cal O}(v_{\rm f}^2)\rho_\gamma$, where
$\rho_\gamma\simeq (\pi^2 g_*/30) T^4$ is the energy density of the electroweak
plasma carrying $g_*$ relativistic degrees of freedom, $T$ is the
temperature and $v_{\rm f}$ is the fluid velocity~\cite{baym,sigl}. 
This means that a magnetic field of size
\begin{equation}
\label{a}
B\sim 0.4\left(\frac{v_{\rm f}}{0.05}\right)\,  T^2\, ,
\end{equation}
can be generated via turbulence.  

One possible mechanism for
generation of magnetic seed fields is by the dipole electromagnetic
charge layer that develops on the surfaces of the bubbles as a
consequence of baryon asymmetry~\cite{baym,sigl}.
The rotation of the dipole charge layer thus sets
up a current in the fluid.  The magnetic field may be generated from
these currents in the bubble walls and then be amplified to the
equipartition value (\ref{a}) by exchange of energy with the turbulent
fluid. Higgs phase gradients can also act as a source for gauge fields
at bubble collisions~\cite{kibble,ae}.  Of course magnetic fields from
the electroweak transition can survive only on scales on which
magnetic diffusion has not had time to wash out the field
correlations. This means that magnetic fields on the bubble radius
scale $R_{\rm b}$ dies off on a time scale $\sim \sigma R_{\rm b}^2$,
where $\sigma\sim 10\, T$~\cite{con} is the conductivity of the
plasma. Being bubble sizes ${\cal O}(10^{-2}-10^{-3})$ of the Hubble
radius at the electroweak phase transition magnetic fields set up at
the electroweak phase transition die away on time scales much larger
than the Hubble time at the phase transition.  Lacking a more precise
calculation about the exact magnitude of the generated magnetic field
at the electroweak phase transition we will parametrize it through the
dimensionless parameter $b$ as
\begin{equation}
B=b\,T^2\, ,
\label{b}
\end{equation}
while from what we have just discussed above values of $b\lesssim 0.4$ seem quite
plausible. Such values are  comfortably smaller than those deduced by imposing 
that the energy density stored in the magnetic field 
at the characteristic scale of production (in our scale the bubble radius) does not appreciably 
alter the  dynamics of primordial nucleosynthesis and the structure of the CMB anisotropies \cite{caprini}.

Now the key point is that sphaleron configurations do
possess a magnetic dipole moment~\cite{Manton}.  In the background of
a magnetic field in the bubble of the broken phase the coupling with
the dipole moment lowers the height of the sphaleron barrier so
that thermal fluctuations are more effective in producing topological
transitions~\cite{CGPR}.  Therefore in order to preserve the baryon asymmetry
within the bubble where a magnetic field is produced by the phase transition 
it is then necessary to require a stronger first order phase
transition than in the case in which one neglects
the presence of the magnetic field. 

The previous comments do apply to any first order phase transition generating EWBG.
However it has been shown that EWBG cannot be realized within the
Standard Model (SM) framework~\cite{AndH, improvement, twoloop, nonpert,CPSM} and it is neither feasible in its Minimal Supersymmetric
extension (MSSM) for arbitrary values of its parameters~\cite{early, mariano1, mariano2}. A particular region in the space of
supersymmetric mass parameters was found in the MSSM, dubbed under the name of light stop
scenario (LSS)~\cite{CQW, Delepine, CK, FL, JoseR, JRB, Carena:1997gx,
Carena:1997ki, CJK, Iiro2, Toni2, Toni3, Worah, Schmidt,
Cline:2000kb, Carena:2000id, Konstandin:2005cd, Cirigliano:2006dg,
qn, chung1, chung2, chung3, window}, where EWBG had
the potential of being successful. In particular the condition
that sphaleron interactions are inhibited in the broken phase required
a sufficiently strong first order phase transition imposing
absolute upper bounds on the lightest CP-even Higgs and right-handed
stop masses, $m_H\lesssim 127$ GeV and $m_{{\widetilde t}_R}\lesssim 120$
GeV~\cite{window}. This is the so-called MSSM baryogenesis window.

In this paper we will consider the EWPT in the MSSM and re-analyze the EWBG
constraints taking into account the magnetic field produced by the phase transition bubbles. It is then clear that the presence of a magnetic field inside the bubble combined with a
non-vanishing magnetic dipole moment of the sphaleron will lower the upper
bound on the lightest CP-even Higgs mass and might close the present EWBG window. However as we have previously stated in the absence of a precise enough calculation we will parametrize the magnetic field by the dimensionless parameter in Eq.~(\ref{b}) so that the results in this paper could be interpreted as upper bounds on the magnitude of the parameter $b$. In this way the MSSM EWBG scenario can only be disproved by future experimental results on the Higgs and right-handed stop masses and/or a more precise theoretical calculation of the produced magnetic field. 

The paper is organized as follows. In Section~\ref{sec:EWBG} we
provide a short summary of the LSS, while in Section~\ref{sec:sphaleron} we briefly discuss the energy of the sphaleron in a magnetic field, deferring the technical details to Appendices~\ref{appendice1} and \ref{appendice2}.  Section~\ref{sec:results}
contains our numerical results and  Section~\ref{sec:conclusions} the conclusions.

%%%%%%%%%%%%%%%%%%%%%%%%%%%%%%%%%%%%%
\section{The Light Stop Scenario and EWBG}
\label{sec:EWBG}
%%%%%%%%%%%%%%%%%%%%%%%%%%%%%%%%%%%%%
%
While in the SM there exists no viable electroweak scale mechanism to explain the
BAU~\cite{AndH, improvement, twoloop, nonpert, CPSM} in the MSSM it
is possible to generate the observed matter-antimatter asymmetry via
EWBG~\cite{CQW, Delepine, CK, FL, JoseR, JRB, Carena:1997gx,
  Carena:1997ki, CJK, Iiro2, Toni2, Toni3, Worah, Schmidt,
  Cline:2000kb, Carena:2000id, Konstandin:2005cd, Cirigliano:2006dg,
  qn, chung1, chung2, chung3, window}. The reason is that the LSS can
overcome the two main problems precluding EWBG to work within the SM:
the impossibility of a strong first order EWPT and the lack of large
CP-violating sources at the EW scale.  Concerning the latter the MSSM
naturally provides new CP-violating interactions. If the 
charginos and neutralinos are light and their mass parameters have non-negligible
relative phases the currents associated to them are sufficient to
produce enough CP violation during the EWPT. On the other hand these
phases affect observables as electric dipole moments (EDM) which are
constrained by experiments. The one-loop contributions to the EDM
may be efficiently suppressed if the first and second generation
scalar particles have a mass equal or larger than
$\mathcal{O}(10)$~TeV. Nevertheless two-loop corrections involving the
charginos and the Higgs field would remain sizable unless the CP-odd
Higgs mass is heavier than ${\mathcal O(1)}$~TeV. Still, even for
large CP-odd Higgs mass, a contribution induced by the SM-like Higgs
cannot be avoided what becomes a LSS prediction that can be tested at
the forthcoming experiments~\cite{EDM}.

Regarding the lack of a strong first order EWPT in the SM, extra
bosons can strengthen the phase transition if their couplings to the
Higgs are sizable and their thermal abundances are not Boltzmann
suppressed. In the MSSM the scalars fulfilling these requirements are
the superpartners of the top quark. In practice only the (mainly)
right-handed stop may be light. In fact the heaviest (mainly)
left-handed stop has to acquire a mass above a few TeV to achieve
agreement with electroweak precision tests and to ensure a
sufficiently heavy SM-like Higgs boson~\cite{qn
%,carena
} compatible with the
LEP bound $m_H>114.4$~GeV~\cite{lep}. On the other hand light gluinos
jeopardize the improvement on the phase transition since their
presence in the plasma increases substantially the thermal mass of the
right-handed stops which then may become Boltzmann suppressed which
 implies that a gluino mass $\gtrsim$500~GeV is
preferred. Finally in order to counteract the remaining thermal mass
contributions to the lightest stop a negative stop square mass term is
required: {\it i.e.}~the right-handed stop is required to be lighter than the top.

In conclusion the spectrum of the LSS at the ${\mathcal
  O}(100)$ GeV scale appears as constituted of the SM spectrum and light charginos, neutralinos
and the right-handed stop. The other
fields, namely gluinos and the remaining scalars, can be decoupled because
the EWBG mechanism is sensitive only to the EW scale. Therefore
the generation of the BAU can be investigated in the low-energy
effective theory where heavy fields are integrated out and their large
radiative corrections are resummed by assuming (for simplicity)  a similar mass $\tm$ for the heavy scalars~\cite{qn}.

At low energy the scalar sector is described by the effective
potential of the SM-like Higgs $H$ and the lightest stop $\widetilde t_R$. At
finite temperature $T$ we approximate it as
\be
\label{effpot} 
V(H,\widetilde t_R,T)= V^{\rm (tree)}(H,\widetilde t_R)+ V^{\rm (rad)}(H,\widetilde
t_R,T)~, 
\ee
where $V^{\rm (rad)}$ includes up to two-loops corrections in the top
Yukawa and strong gauge couplings~\footnote{To calculate
  $V(H,\widetilde t_R,T)$ we choose the low energy spectrum obtained
  in Ref.~\cite{qn,window} to which we refer for the explicit
  expressions of $V^{\rm (tree)}(H,\widetilde t_R)$ and
  $V^{\rm (rad)}(H,\widetilde t_R,T)$.}. In fact the potential
$V(H,\widetilde t_R,T)$ is required to analyze the phase
transition. By using the bounce method one can determine the
nucleation temperature $T_n$~\cite{Linde, thomas} below which the
bubbles containing the electroweak breaking phase can form, expand and
then collide to each other. This process goes on till the temperature
$T_f ~(<T_n)$ when the Universe is completely filled of the
electroweak broken phase and the transition is ended.

The successful production of the BAU can be synthetized into three main
conditions. First, bubbles of electroweak broken phase must nucleate
and fill the Universe. Second, the baryon asymmetry must be produced
and injected into the bubbles of the broken phase. Third, inside the
bubbles $SU(2)_L$ sphalerons must be out of equilibrium not to wash
out the baryon asymmetry. In particular, in the limit of vanishing magnetic field $B\to0$ three main implications follow~\cite{window}:
\begin{enumerate}
\item Since the square mass term of the lightest stop is negative
  $V(H,\widetilde t_R,T=0)$ contains a color-breaking minimum along
  the direction ($H=0,\widetilde t_R$) besides the standard
  minimum at ($H=(0,v/\sqrt{2})^T,\widetilde t_R=
    0$)~\footnote{We use the convention $v\equiv v(T=0)= 246.2 \GeV$
    with $v(T)$ being the Vacuum Expectation Value (VEV) of the Higgs in the
    EW-breaking minimum at temperature $T$.}. Therefore it is
  possible that portions of the Universe decay into the color breaking
  phase from where they cannot escape. To avoid such a problem the
  transition towards the EW breaking phase must end before any
  tunneling into the color breaking minimum is allowed. This
  requirement is roughly guaranteed by the condition $T_c\gtrsim
  T_{\rm col}+1.6\,\GeV$, where $T_c$ and $T_{\rm col}$ are the
  temperatures at which the minimum of the unbroken phase is
  degenerate with the electroweak and color breaking ones,
  respectively.
\item The amount of baryon asymmetry injected into the bubbles is
  enhanced nearby a resonance region occurring for degenerate gaugino
  and Higgsino
  masses~\cite{Carena:2000id,Konstandin:2005cd,Cirigliano:2006dg, qn,
    chung1, chung2, chung3}. In such a case, requiring enough BAU
  implies a bound on the ratio between the VEVs of the up and down
  Higgses of the MSSM, precisely $\tbe\lesssim 15$.  The same bound
  provides suppression of the EDM within the experimental constraints.
\item The condition that sphaleron transitions are inhibited in the
  broken phase requires a sufficiently strong first order phase
  transition. Quantitatively this depends on the relation between
  $V(H,\widetilde t_R=0,T)$ at $T\lesssim T_n$ and the $SU(2)_L$
  sphaleron rate. This requirement imposes absolute upper bounds on
  the lightest SM-like Higgs and right-handed stop masses as $m_H\lesssim
  127$ GeV and $m_{{\widetilde t}_R}\lesssim 120$ GeV \cite{window}.
  These extreme values are obtained for $\tm $ beyond the PeV scale while 
  smaller values of $\widetilde{m}$ provide smaller upper
  bounds. In particular for  $\widetilde m\lesssim$ 6 TeV they lie under the experimental constraints: $m_H>114.4\,
  \GeV$ \cite{lep} and $m_{\widetilde t_R}\gtrsim 95\,\GeV$~\cite{cdf}.
\end{enumerate}
However one must be worried about the formation of the magnetic field that
may substantially modify these conclusions.

As explained in the introduction the dynamics of the phase transition 
may  enhance the seeds of the magnetic fields to sizeable values~\cite{baym,sigl}.
Its magnitude is small at bubble nucleation so  that nucleation properties are  not altered by it. 
Instead the magnetic
field may become sizable  during the phase transition  and we consider it as a
homogeneous background field $B$ in the whole interior of the bubble.
A large $B$ affects the effective potential $V$~\cite{Navarro}
but the effect can be neglected because it arises
beyond the approximation we use to calculate $V(H,\widetilde t_R,T)$
of Eq.~(\ref{effpot}).  
In the approximation we are using, where gauge boson loops play a subleading role, the gauge dependence of the effective potential \cite{gaugedep} would affect only mildly the sphaleron energy and its effects should be comparable to other subleading effects we have not considered.
Finally we assume that $B$ does not modify
significantly the diffusion processes injecting the baryon asymmetry
into the interior of the bubble. For this
reason we conclude that the presence of the magnetic field $B$
threatens EWBG in the LSS through alterations of the above third
implication. In fact, as we will explain in detail in the next
section, the magnetic field changes the standard link between
$SU(2)_L$ sphaleron rate and $V(H,\widetilde t_R,T)$ thus reducing the
parameter space where the phase transition is strong.

%%%%%%%%%%%%%%%%%%%%%%%%%%%%%%%%%%%%%
\section{Sphaleron in a magnetic field}
\label{sec:sphaleron}
%%%%%%%%%%%%%%%%%%%%%%%%%%%%%%%%%%%%%

\noindent
For vanishing weak mixing angle, $\theta_{\rm w}=0$, the sphaleron
solution is spherically symmetric and does not develop any magnetic
dipole moment.  For $\theta_{\rm w}\neq 0$ the $U(1)_Y$ gauge field
$a_\mu$ is excited and the spherical symmetry reduces to an axial
symmetry.  A magnetic dipole moment is present and can be computed at
the lowest order in $\theta_{\rm w}$ using the sphaleron solutions
obtained for $\theta_{\rm w}=0$ \cite{Manton}.  Because of the very
weak dependence on $\theta_{\rm w}$ the discrepancy with respect to
computing the dipole moment with the sphaleron solutions at
$\sin\theta_{\rm w}\simeq 0.48$ is less than 1\% \cite{Laterveer}.  A
description of the sphaleron solutions and their energy can be found
in Appendix~\ref{appendice1}, while the sphaleron magnetic dipole
moment is discussed in Appendix~\ref{appendice2}.

In the presence of a magnetic field, Eq.~(\ref{b}), inside the bubbles the
sphaleron energy changes due to the interaction of the magnetic field
with the sphaleron magnetic dipole moment
 \be E_{\rm sph}(T,B)=E_{\rm
  sph}(T) + E_{\rm dipole}(T,B)\,,
\ee
where $E_{\rm sph}(T)$ is the sphaleron energy computed without
magnetic field, but taking into account both the temperature
dependence of the Higgs potential and the sphaleron solutions for
$\theta_{\rm w}\neq 0$ [see Eq.~(\ref{Esphaleron})].
The dipole energy to leading order in $\theta_{\rm w}$ is computed
using Eqs.~(\ref{Edipole})-(\ref{mu}) 
\be 
E_{\rm  sph}^{(1)}(T,B)=E_{\rm sph}(T) + E_{\rm dipole}^{(1)}(T,B)=E_{\rm
  sph}(T)-\mu(T) B\,.
\label{EsphB}
\ee
Beyond leading order in $\theta_{\rm w}$ a non-linear dependence
of the sphaleron energy on the magnetic field also arises. However
for the range of magnetic fields that are relevant for our analysis
the corrections to the linear approximation (\ref{EsphB}) are less
than 5\%, as discussed in Ref.~\cite{CGPR}.

\begin{figure}[t]
\centering
   \includegraphics[height=5cm, width=7.5cm]{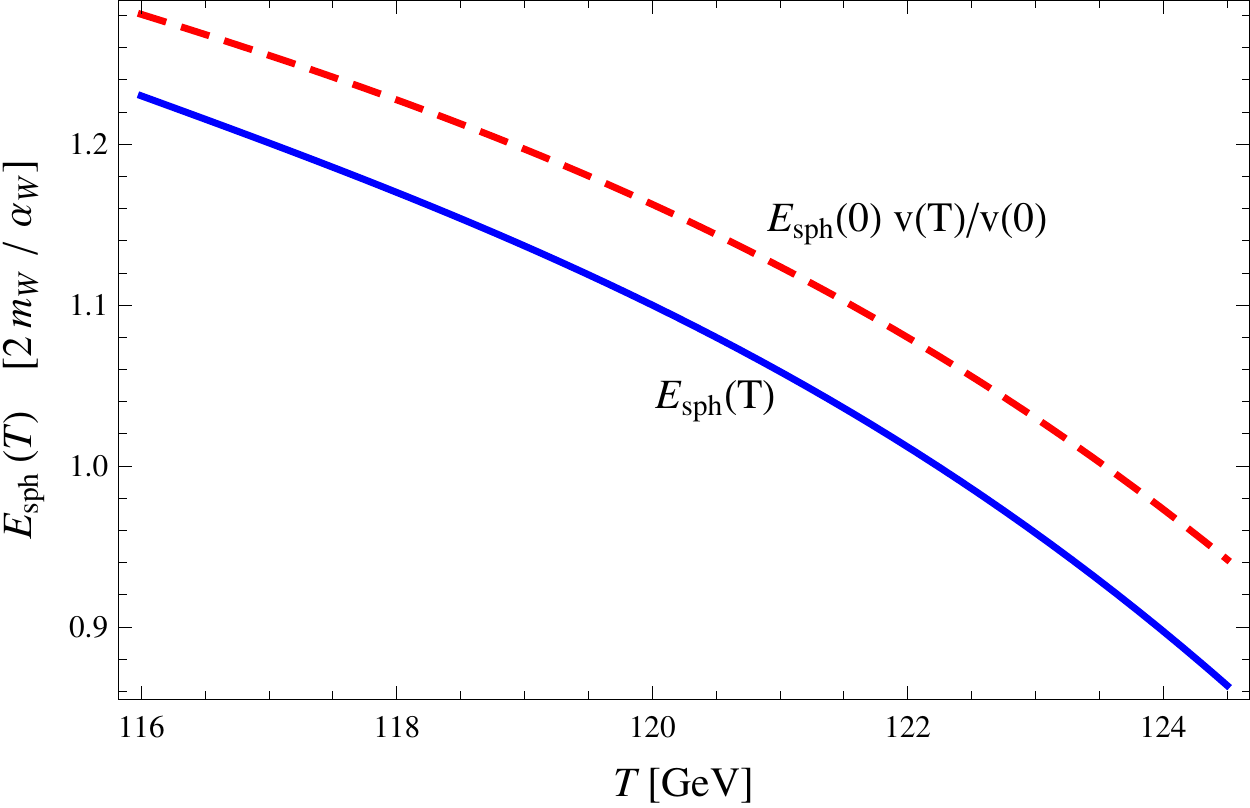} \quad
   \includegraphics[height=5cm, width=7.5cm]{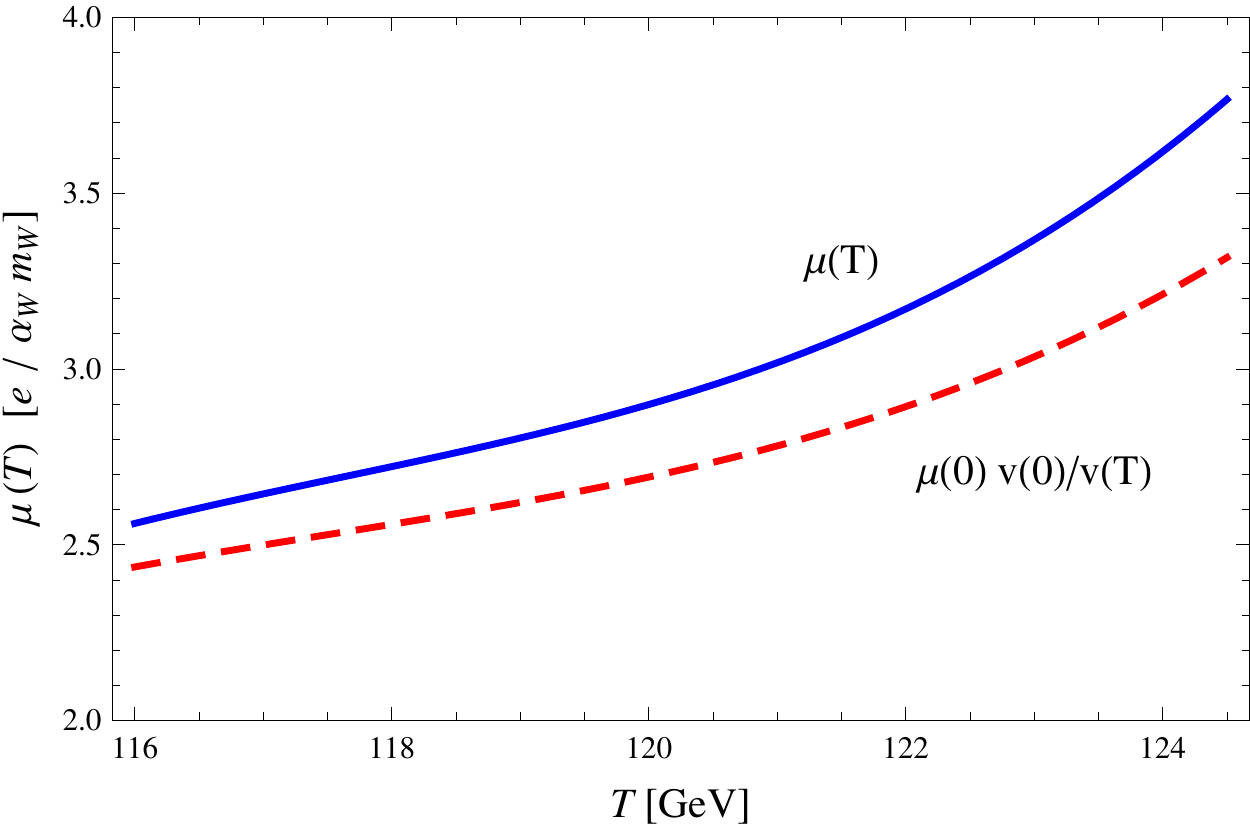}\\
   \includegraphics[height=5cm, width=7.5cm]{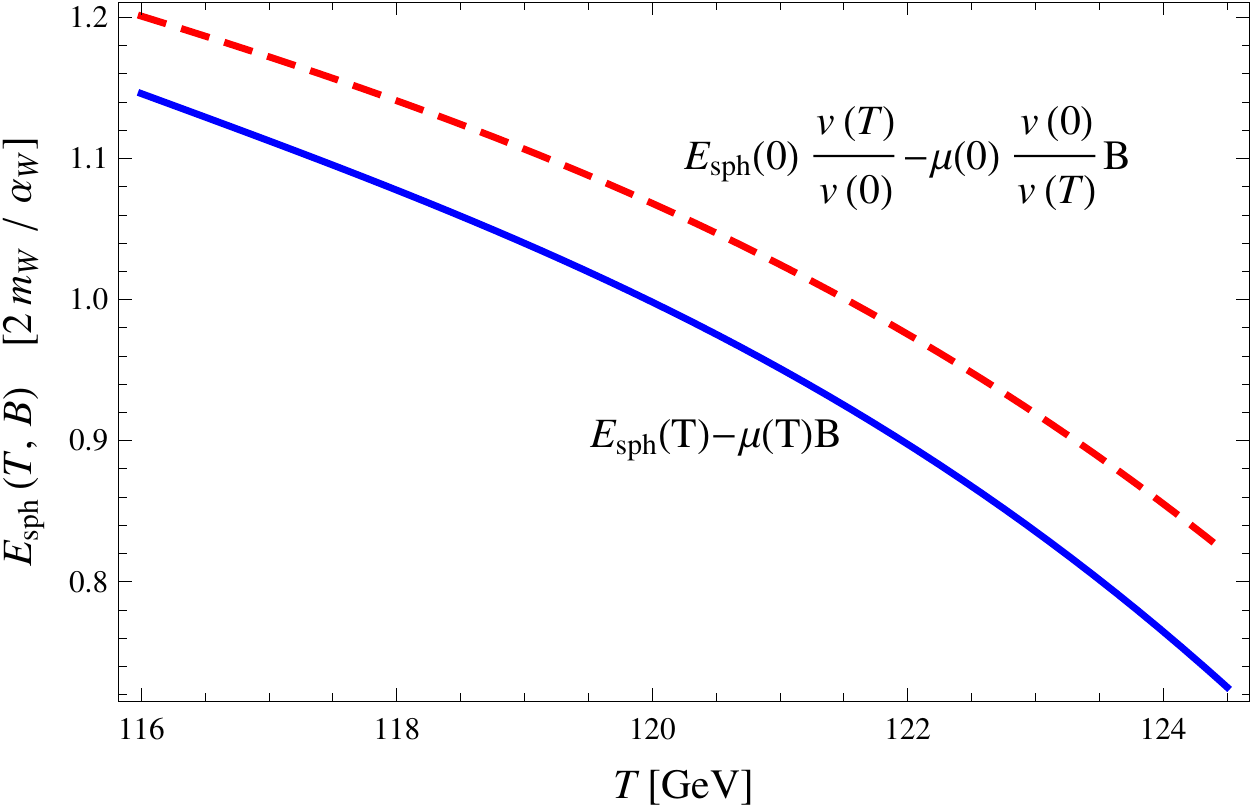} 
 \caption{{\small \emph{The temperature dependence 
 of the sphaleron energy (with and without magnetic field)  and the magnetic dipole moment.
\emph{Top left panel:}  sphaleron energy without magnetic field;
\emph{Top right panel:}  sphaleron magnetic dipole moment;
\emph{Bottom panel:} sphaleron energy for $B=0.1\,T^2$.
We have fixed  $m_H=120.2 \GeV $ and $\widetilde m=10^3 \TeV$.
We show both the results of numerical calculations using the Higgs potentials at nonzero temperatures (blue solid line)
and the result of a simple scaling relation (red dashed line).
	}}}
 \label{fig:scaling}
\end{figure}

To capture most of the temperature dependence of the sphaleron
energy there exists a very common approximation which avoids
resorting to solve for the sphaleron functions at $T\neq 0$.  It
consists in assuming that the whole temperature dependence is encoded
in the expectation value $v(T)$. For vanishing magnetic field the
scaling law is
\be
E^{\rm scaling}_{\rm sph}(T)=E_{\rm sph}(0) {v(T)\over v(0)}\,,
\label{scaling}
\ee 
which overestimates the correct energy by about 10\%, as shown in
Ref.~\cite{sphaleronMSSM} (top left panel of
Fig.~\ref{fig:scaling}). On the other hand the dipole moment in
Eq.~(\ref{mu}) scales with the inverse of $v(T)$ and this is accurate
to better than 15\% (top right panel of Fig.~\ref{fig:scaling}).  The
scaling law for the total energy in presence of a magnetic field,
Eq.~(\ref{EsphB}), is given by the combination of the two scalings
(bottom panel of Fig.~\ref{fig:scaling}).  In this paper we have not
made use of the scaling law but instead directly computed the
sphaleron solutions and the corresponding energy at non-zero
temperature.

The condition for sphaleron transitions going out of equilibrium and
not washing out the baryon asymmetry is
\cite{sphaleron_rate}~\footnote{One could prefer carrying out the
  analysis using a condition on $E_{\rm sph}(T)/T$ at $T=T_c$ rather
  than $T=T_n$. We find that $E_{\rm sph}(T_c)$ is lower than $E_{\rm
    sph}(T_n)$ by about 15\%, and the analogue of the bound
  (\ref{ETn}) becomes $E_{\rm sph}(T_c)/T_c\gtrsim 29$. If one adopted the
  scaling approximation (\ref{scaling}), the $\mathcal{O}(10\%)$
  overestimate of the sphaleron energy would partially cancel the
  mismatch between the energies computed at $T_c$ and at $T_n$.}
\be
{E_{\rm sph}(T_n,B)\over T_n}\gtrsim 35\,,
\label{ETn}
\ee
where for simplicity we assume the magnetic field to become constant
and sizable just after the bubble nucleations \footnote{Considering
  the $B$ field to arise at a temperature $T_B$ as low as $T_f$ would tend to
  relax the bounds we will provide. However it seems realistic to
  consider that the main collisions occur at $T\simeq T_n$ and therefore we believe our final conclusion should be conservative. However lacking a detailed calculation on the generation of the magnetic field in the LSS first order phase transition we have not included the uncertainty in the determination of $T_B$ in the plots.}.
Implicitly, Eq.~(\ref{ETn}) provides a constraint on $V(H,\widetilde
t_R,T_n)$ and for $B=0$ it is roughly satisfied by the upper bounds
$m_H\lesssim 127\,\GeV$ and $m_{{\widetilde t}_R}\lesssim 120\,\GeV$
\cite{window}. Instead, for $B\neq 0$, these upper bounds become more
stringent because of the negative contribution $E_{\rm dipole}^{(1)}$
in Eq.~(\ref{EsphB}). The numerical analysis will be done in the next section.

\begin{figure}[t]
   \includegraphics[height=6cm, width=8cm]{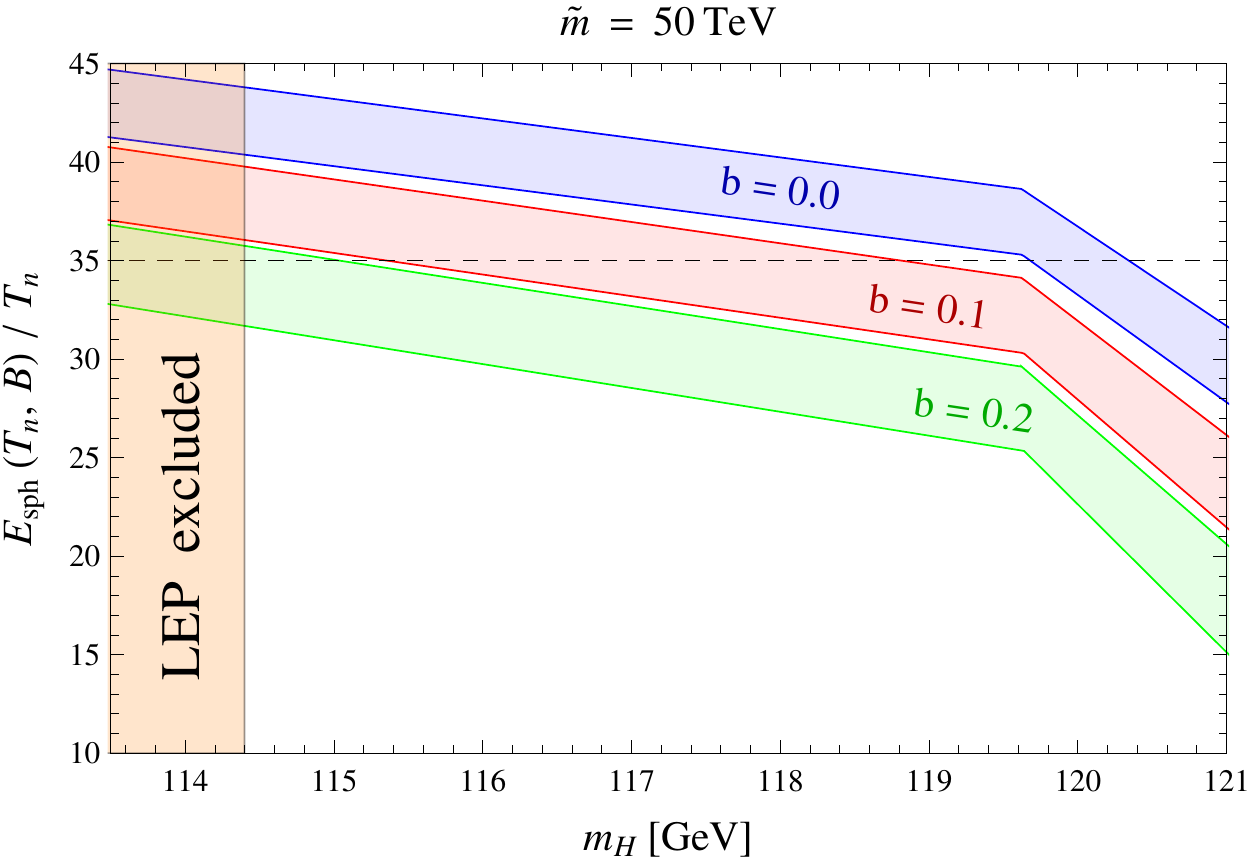} \quad
   \includegraphics[height=6cm, width=8cm]{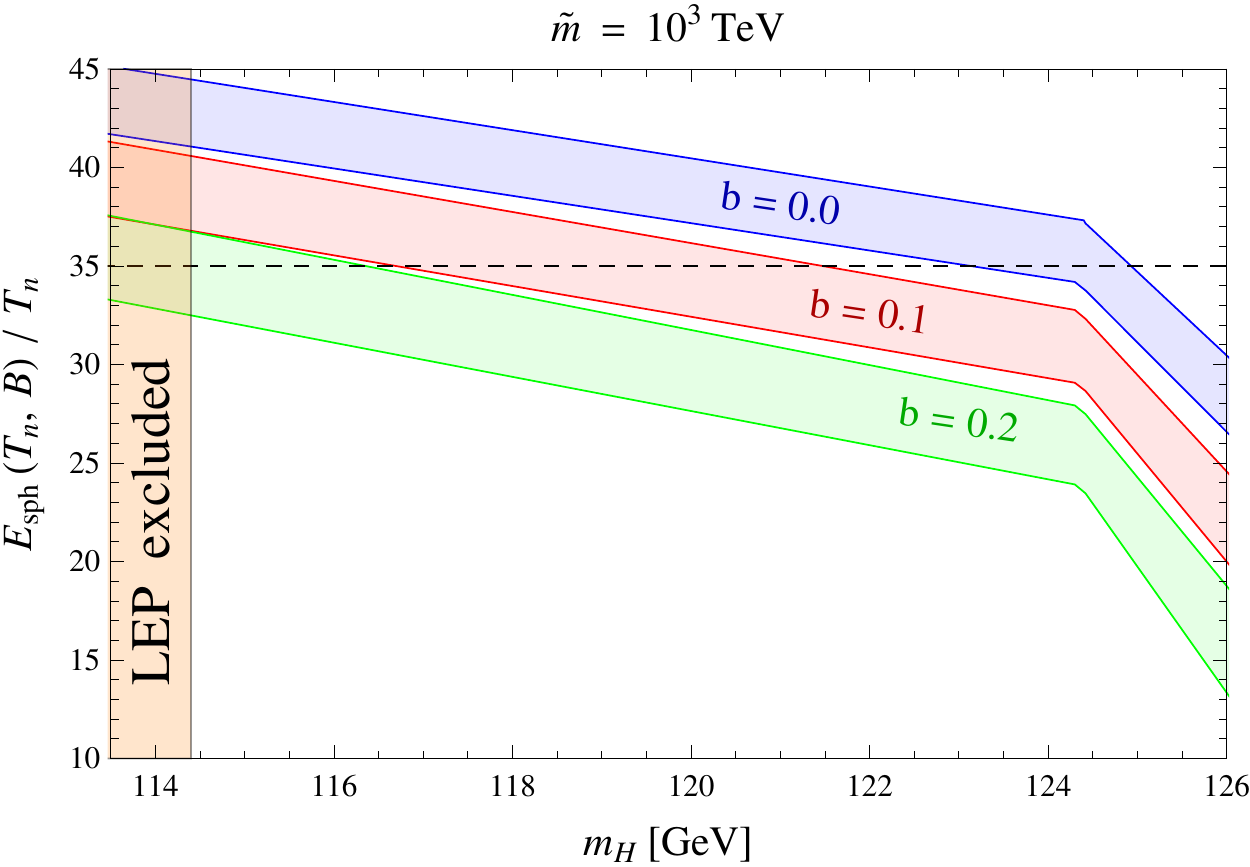}
   \caption{{\small \emph{The maximal sphaleron energy $E(T_n, B)/T_n$
         achieved for a Higgs mass $m_H$ evaluated with $\widetilde
         m=50 $ TeV \emph{(left panel)} and $\widetilde m=10^3 \TeV $
         \emph{(right panel)}.  The horizontal dashed line corresponds
         to the requirement $E(T_n)/T_n\gtrsim 35$.  The bands
         correspond to the uncertainty on the location of $T_n$ in the
         interval $[T_c-3.5 \GeV, T_c-2 \GeV]$.  The upper lines
         correspond to $T_n=T_c-3.5 \GeV$.  Different bands correspond
         to different values of $b=B/T_n^2= 0.0, 0.1, 0.2$.  }}}
 \label{fig:energy}
\end{figure}

%%%%%%%%%%%%%%%%%%%%%%%%%%%%%%%%%%%%%
\section{Numerical results}
\label{sec:results}
%%%%%%%%%%%%%%%%%%%%%%%%%%%%%%%%%%%%%

\noindent
We have considered a sample of points of the parameter space where we
calculate $V(H,\widetilde t_R,T)$ of Eq.~(\ref{effpot}) and we use it
to determine $T_n$ by the bounce method~\cite{Linde, thomas}. We
observe that in the subset of points fulfilling the condition
(\ref{ETn}) with $B/T_n^2\le 0.2$ we get $T_c-3.5 \GeV<T_n< T_c-2
\GeV$ which we will translate into an error in the determination of
$T_n$. Then we perform a wide scan in the LSS parameter space and at
each point we calculate the effective potential of Eq.~(\ref{effpot})
and the correspondingly quantities $m_H,m_{\tilde t_R},T_c$. From
$T_c$ we determine $T_n$ with the error $[T_c-3.5 \GeV, T_c-2 \GeV$]
and subsequently $E_{\rm sph}(T_n)$.

\begin{figure}[t]
\centering
  \includegraphics[height=6cm, width=8cm]{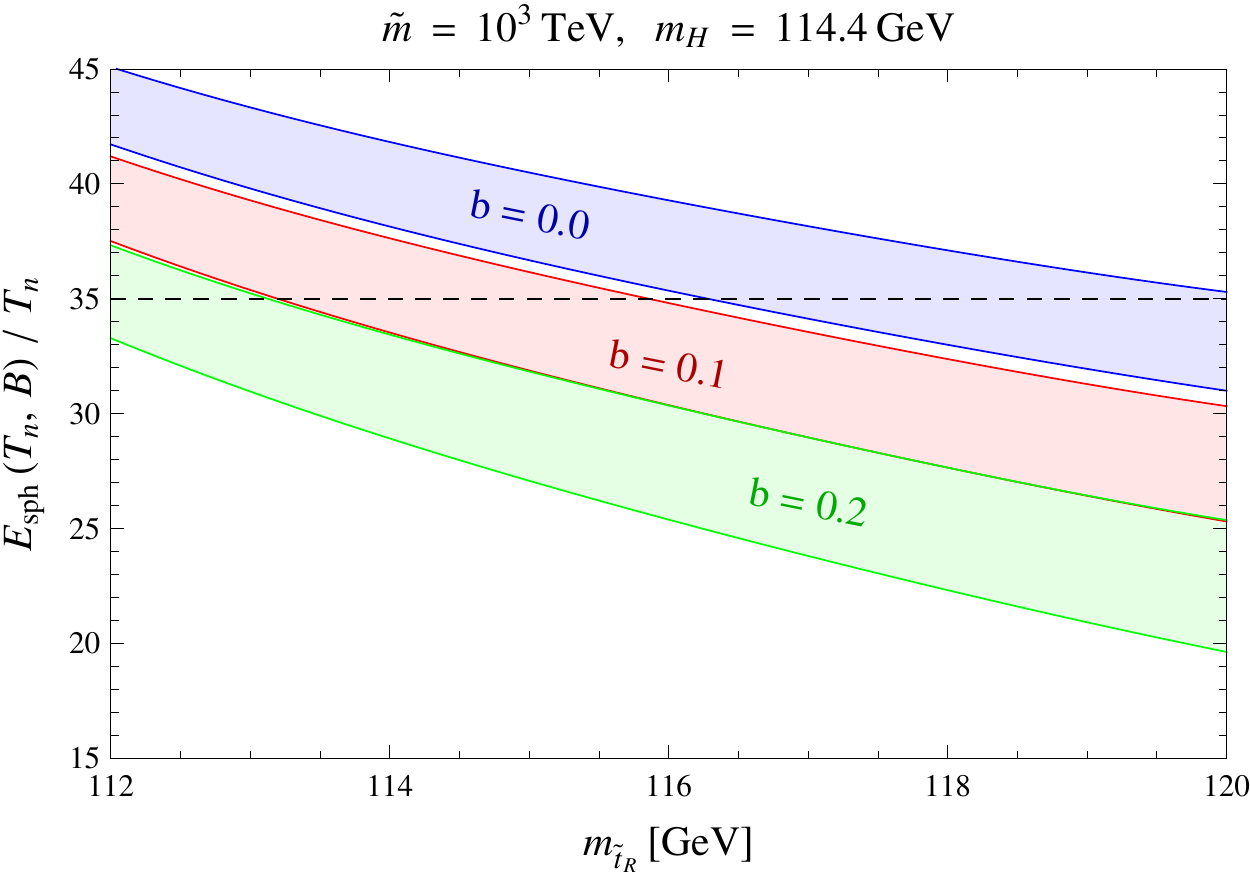}
    \includegraphics[height=6cm, width=8cm]{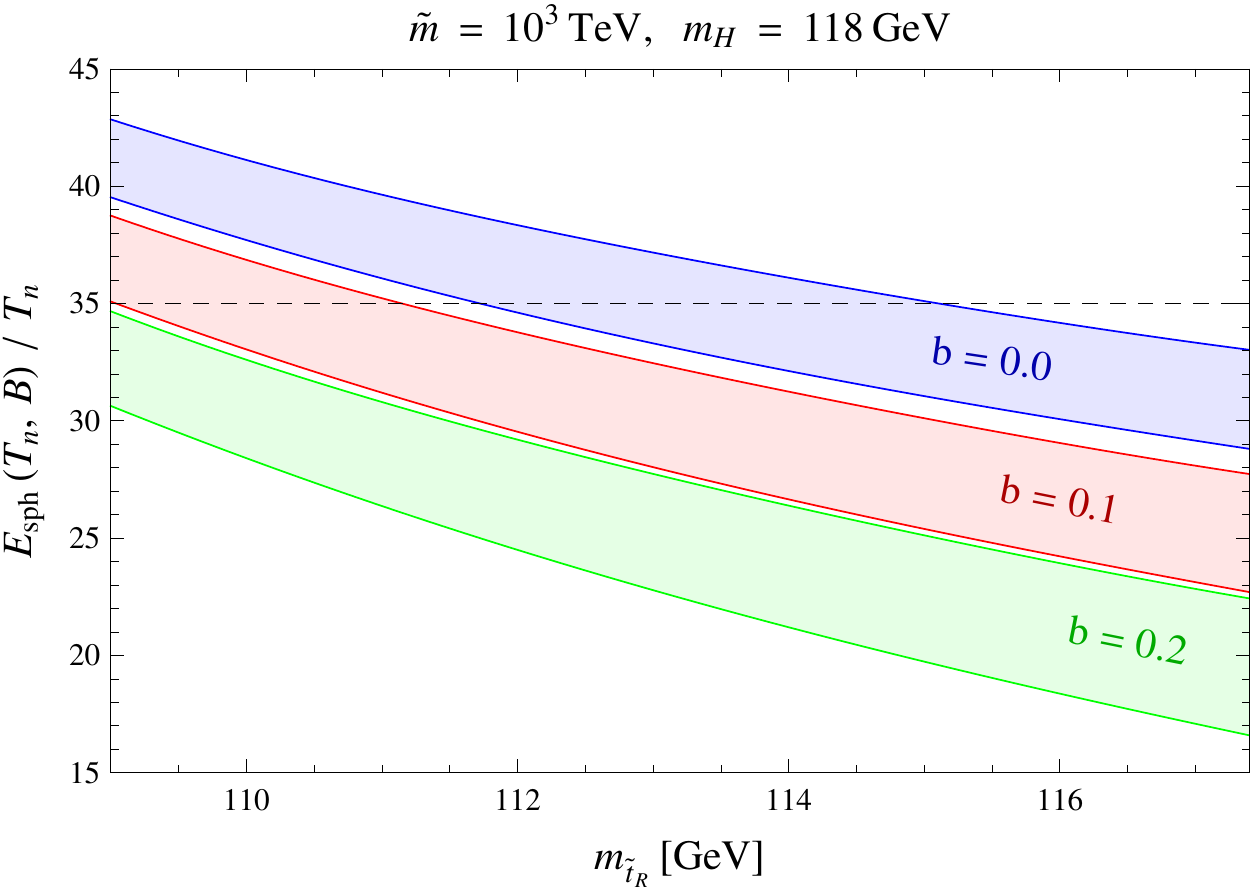}
    \caption{{\small \emph{ The sphaleron energy $E(T_n, B)/T_n$ as a
          function of the maximal right-handed stop mass
          $m_{\widetilde t_R}$ allowed for $\widetilde m=10^3 \TeV$
          and $m_H=114.4 \GeV $ \emph{(left panel)} or $m_H=118 \GeV$
          \emph{(right panel)}.  The horizontal dashed line
          corresponds to the requirement $E(T_n)/T_n\gtrsim 35$.  The
          bands correspond to the uncertainty on the location of $T_n$
          in the interval $[T_c-3.5 \GeV, T_c-2 \GeV]$.  The upper
          lines correspond to $T_n=T_c-3.5 \GeV$.  Different bands
          correspond to different values of $b=B/T_n^2= 0.0, 0.1,
          0.2$.  }}}
 \label{fig:mstop}
\end{figure}

Using this procedure we fix $\tm$ and look for the maximal Higgs mass
achieving a fixed value of $E_{\rm sph}(T_n,B)/T_n$. 
The result is shown in Fig.~\ref{fig:energy} for $\widetilde m=50\TeV$ and $10^3 \TeV$, 
and $b=B/T_n^2=0.0,0.1,0.2$. The value of the stop mass is 
conveniently chosen to maximize the Higgs mass. 
The slope of the curves changes, giving rise to the kinks observed in the figure,
when the stop mass drops below the experimental bound 
$m_{\tilde t_R}\geq  95\,\GeV$ \cite{cdf} and therefore one needs to set the stop
mass to its lower limit.
 The bands correspond to the uncertainty on the determination of $T_n$. 
Below the horizontal dashed line, Eq.~(\ref{ETn}) is not satisfied and
EWBG in the LSS does not produce enough baryon asymmetry.
We similarly repeat the scan by fixing $m_H$ and looking for the
maximal allowed stop mass consistent with a given value of $E_{\rm
  sph}(T_n)/T_n$. The outcome is presented in Fig.~\ref{fig:mstop} for
$\tilde m=10^3 \TeV$ and $m_H=114.4\ \GeV$ (left panel) and
$m_H=118.0\ \GeV$ (right panel). The maximal stop mass is now attained
for the minimal Higgs mass and it gets lower when a magnetic field is
present, although remaining above the experimental limit.

\begin{figure}[t]
\centering
   \includegraphics[scale=0.8]{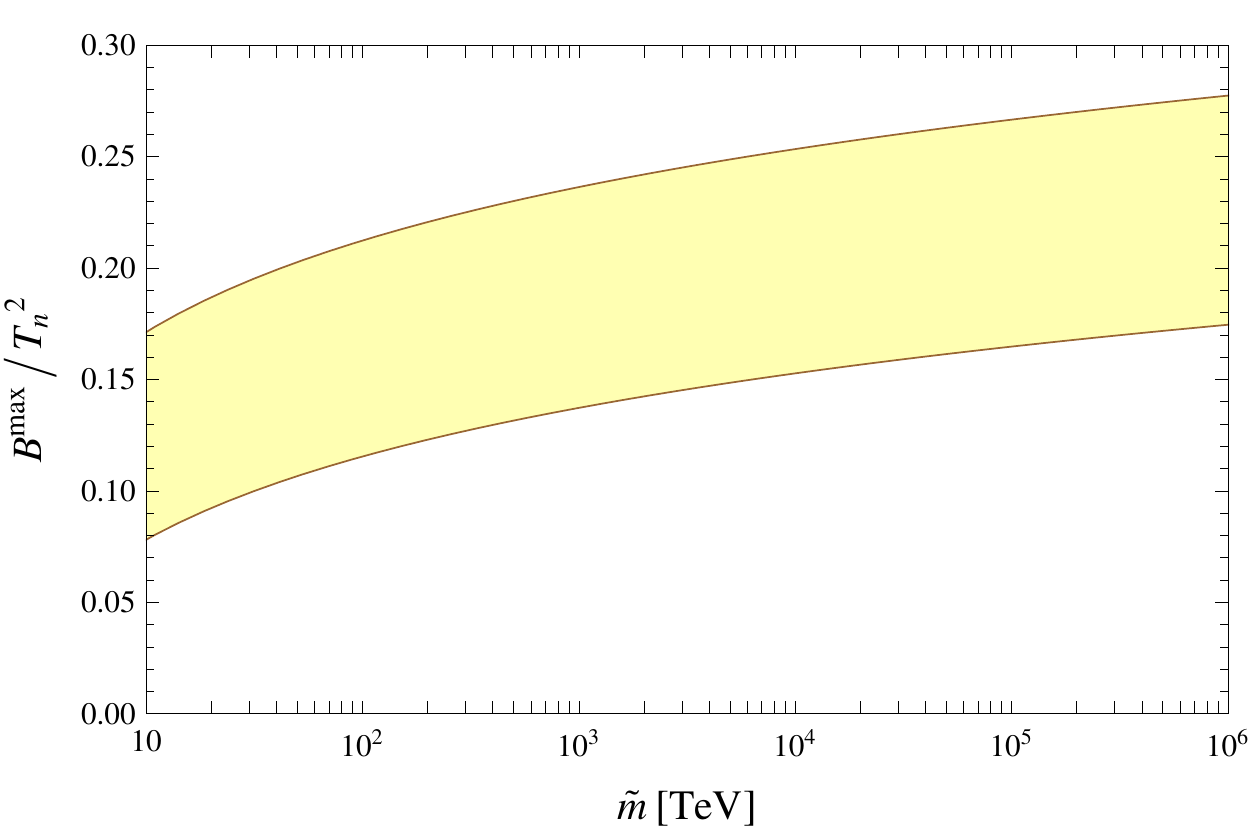}
   \caption{{\small \emph{The values of the magnetic field for which
         the maximal $m_H$ is at the experimental bound.  The band
         corresponds to the uncertainty on the location of $T_n$ in
         the interval $[T_c-3.5 \GeV, T_c-2 \GeV]$.  The upper line
         corresponds to $T_n=T_c-3.5 \GeV$.  }}}
 \label{fig:bmax}
\end{figure}

The previous results can be translated into absolute upper bounds on
the produced magnetic fields which are consistent with the requirement
of EWBG in the MSSM. In Fig.~\ref{fig:bmax} we show the values of the
constant magnetic field generated at $T=T_n$ which pushes the maximum
value of the Higgs mass required by successful EWBG in the LSS down to
the present experimental bound.  Again the band in Fig.~\ref{fig:bmax}
corresponds to the uncertainty on $T_n$.  We see that even moderate
values of magnetic fields generated during the EW phase transition
could close the EWBG window.

%%%%%%%%%%%%%%%%%%%%%%%%%%%%%%%%%%%%%
\section{Conclusions}
\label{sec:conclusions}
%%%%%%%%%%%%%%%%%%%%%%%%%%%%%%%%%%%%%

In this paper we have pointed out that the sphaleron
magnetic dipole moment couples to the magnetic field generated during
a first order EWPT. This has the effect of lowering the sphaleron
energy in the broken phase and consequently the baryon asymmetry is washed out more
easily.  We have not attempted to compute precisely the magnetic field
generated at the phase transition,  but rather we have considered it as
a free parameter within a plausible range.  In particular we have focused on the MSSM in the
most favourable situation for EWBG, the light stop scenario, and
explored the dependence of the sphaleron energy on the parameters of
the model and on the magnetic field.  We have shown that it is
possible to have the baryon asymmetry preserved even in presence of a
magnetic field by lowering the Higgs mass and/or the stop mass, or
increasing the soft scalar mass $\widetilde m$. However even for
moderate values of the magnetic field the required Higgs mass can fall
below the present experimental bound and the window for EWBG in the
MSSM gets closed.

The main conclusions of this paper are twofold.  On the one hand our
calculation cries out  for an accurate determination of $B$
and its evolution during the phase transition in order to confirm 
if the magnetic field produced by the electroweak phase transition
 endangers the EWBG in the LSS, as our  analysis seems to indicate.
On the other hand, the magnetic field produced by phase transition
bubbles  threatens any model where the BAU is generated
by a first order phase transition. In principle, any model of EWBG
where the phase transition (evaluated at zero magnetic field) is never extremely
strong should be jeopardized by the presence of a magnetic field.

\section*{Acknowledgments}
We are grateful to the organizers of the Workshop ``Electroweak
Baryogenesis in the Era of the LHC'', at the Weizmann Institute, for
creating a very stimulating and pleasant environment, where the idea
of this paper originated. We also thank Chiara Caprini for useful
discussions about the limits on magnetic fields generated at the electroweak
phase transition.  The work is supported by the
Swiss National Science Foundation under contract 200021-125237; by the
Spanish Consolider-Ingenio 2010 Programme CPAN (CSD2007-00042); by
CICYT, Spain, under contract FPA 2008-01430; by the Belgian IISN
convention 4.4514.08 of FNRS.

\appendix

%%%%%%%%%%%%%%%%%%%%%%%%%%%%%%%%%%%%%
\section{Sphaleron solutions}
\label{appendice1}
%%%%%%%%%%%%%%%%%%%%%%%%%%%%%%%%%%%%%

Let us consider the classical finite energy configurations of the
bosonic fields of the electroweak sector of the SM, in a gauge where
the time components of the gauge fields are set to zero~\cite{Manton,
  DHN}.  The classical energy functional over configuration space, at
temperature $T$, is 
\be E_{\rm sph}(T)=\int d^3 x\left[ {1\over
    4}F^a_{ij}F^a_{ij}+{1\over 4}f_{ij}f_{ij}+ (D_i H)^\dag (D_i H)+
  V(H, T)\right]\,, 
  \ee 
  where 
  \bea
F_{ij}^a&=& \partial_i W_j^a-\partial_jW_i^a+g \epsilon^{abc}W_i^b W_j^c \,,\\
f_{ij}&=& \partial_i a_j-\partial_j a_i \,,\\
D_i H&=& \partial_i H-{1\over 2}i g\sigma^a W_i^a H-{1\over 2}i g' a_i
H\,, 
\eea 
$W^a_\mu$ $(a=1,2,3)$ and $a_\mu$ are the $SU(2)$ and $U(1)$ gauge
fields, respectively, and $H$ is the Higgs doublet. The gauge
couplings of $SU(2)_L$ and $U(1)_Y$ are $g$ and $g'$, respectively,
and the weak mixing angle is defined by $\tan\theta_{\rm w}=g'/g$. The
potential $V$ is defined as $V(H,T)\equiv V(H,\widetilde t_R={
  0},T)$ as expressed in Eq.~(\ref{effpot}) and in
Refs.~\cite{qn,window}.

\subsection{Case $\theta_{\rm w}=0$}
\label{app:nothetaw}

In this limit the $U(1)_Y$ gauge field $a_i$ decouples and it may be set to zero.
The sphaleron is a spherically symmetric configuration of gauge and Higgs fields.
Let us consider the ansatz for the fields~\cite{Manton, Laterveer}
\bea
g W_i^a \sigma^a dx^i&=&(1-f(\xi)) F_a \sigma^a \,, \\
H&=& {v(T)\over \sqrt{2}}\left(\begin{matrix} 0\\ h(\xi)\end{matrix}\right)\,,
\eea
in terms of two radial functions $f(\xi)$ and
$h(\xi)$ where the dimensionless distance $\xi\equiv g v r$ has been
introduced, $\sigma^a (a=1,2,3)$ are the Pauli matrices and $F_a$ are
the 1-forms~\cite{Laterveer}
\bea
F_1&=& -2\sin\phi d\theta-\sin2\theta \cos\phi d\phi \,,\\
F_2&=& -2\cos\phi d\theta+\sin2\theta \sin\phi d\phi \,,\\
F_3&=& 2\sin^2\theta  d\phi \,.
\eea
After the redefinition
\be
\widetilde V(h,T)\equiv
\left.V(H,T)\right|_{H\to(0,v(T)h/\sqrt{2})^T}~,
\ee
the sphaleron energy is a function of the radial functions
\be
E^{(\theta_{\rm w}=0)}_{\rm sph}(T,B=0)= {4\pi v(T)\over g}\int_0^\infty d\xi \left[4{f'}^2+{8\over \xi^2} f^2(1-f)^2+
{1\over 2}\xi^2 {h'}^2+h^2(1-f)^2+ \xi^2 {\widetilde V(h,T)\over g^2 v^4(T) }\right]
\ee
which is minimized by the solution of the  variational field equations (prime denotes derivative
with respect to $\xi$)
\bea
&& f''-{2\over \xi^2}f(1-f)(1-2f) +{1\over 4}h^2 (1-f)    = 0  
\label{fpp}\,,\\
&& h''+{2\over \xi}h'-{2\over \xi^2}h(1-f)^2 
-{1\over g^2 v^(T)}{\partial \widetilde V(h,T)\over \partial h}    = 0\,.
\label{hpp}
\eea
with boundary conditions $f(0)=h(0)=0,\ f( \infty)=h(\infty)=1$.

\subsection{Case $\theta_{\rm w}\neq 0$}
\label{app:thetaw}

When $\theta_{\rm w}\neq 0$ the sphaleron is not spherically symmetric but only axisymmetric.
The most general ansatz requires seven independent functions of the spherical coordinates
$r$ and $\theta$~\cite{seven_functions}.
However since the dependence on $\theta$ is very mild an excellent approximation to the  
exact solution is provided
by an ansatz in terms of only four scalar functions of $r$~\cite{Laterveer}
\bea
g' a_i \, dx^i&=&(1-f_0(\xi)) F_3\,, \\
g W_i^a \sigma^a dx^i&=&(1-f(\xi)) \left(F_1 \sigma^1 +F_2 \sigma^2\right)+(1-f_3(\xi)) F_3\sigma^3 \,, \\
H&=& {v(T)\over \sqrt{2}}\left(\begin{matrix} 0\\ h(\xi)\end{matrix}\right)\,.
\eea
Now the sphaleron energy reads
\bea
E_{\rm sph}(T)&=& {4\pi v(T)\over g}\int_0^\infty d\xi \left\{
{8\over 3}{f'}^2+{4\over 3}{f_3'}^2+ 
{8\over \xi^2}\left[ {2\over 3} f_3^2(1-f)^2+{1\over 3}\left(f(1-f)+f-f_3\right)^2 \right]
\right.\nn \\
&&
+{4\over 3}\left({g\over g'}\right)^2\left[
{f_0'}^2+{2\over \xi^2}(1-f_0)^2
\right]
+{1\over 2}\xi^2{h'}^2
+h^2\left[{1\over 3}(f_0-f_3)^2+{2\over 3}(1-f)^2\right] \nn\\
&&\left.
+ \xi^2 {\widetilde V(h,T)\over g^2 v(T)^4 }\right\} \,,
\label{Esphaleron}
\eea
which is minimized by the solutions of the variational equations
\bea
&&f''+{2\over \xi^2}(1-f)\left[f(f-2)+f_3(1+f_3)\right] +{1\over 4}h^2 (1-f)    = 0  \,,\\
&&f_3''-{2\over \xi^2}\left[3f_3+f(f-2)(1+2f_3)\right]+{1\over 4}h^2 (f_0-f_3)    = 0  \,,\\
&&f_0''+{2\over \xi^2}(1-f_0)-{{g'}^2\over 4g^2 }h^2 (f_0-f_3)    = 0  \,,\\
&&h'' +{2\over \xi}h'-{2\over 3 \xi^2}h\left[2(1-f)^2+(f_0-f_3)^2 \right]
-{1\over g^2 v(T)^4}{\partial \widetilde V(h,T)\over \partial h}   =0\,.
\eea
with boundary conditions $f(0)=f_3(0)=h(0)=0,\ f_0(0)=1$ and
$f( \infty)=f_3(\infty)=f_0(\infty)=h(\infty)=1$.
For $\theta_{\rm w}\to 0$ one recovers the solutions of the previous subsection:
$f_0(\xi)\to 1, f_3(\xi)\to f(\xi)$.

%%%%%%%%%%%%%%%%%%%%%%%%%%%%%%%%%%%%%
\section{Sphaleron magnetic dipole moment}
\label{appendice2}
%%%%%%%%%%%%%%%%%%%%%%%%%%%%%%%%%%%%%

A magnetic dipole moment for the sphaleron arises when $\theta_{\rm w}\neq 0$~\cite{Manton}.
The $U(1)_Y$ gauge field $a_i$  is not decoupled, as in the case $\theta_{\rm w}=0$,
and it cannot be set to zero because it is sourced by the current
\be
J_i=-{1\over 2}i g' \left[H^\dag D_i H-(D_i H)^\dag H\right]\,,
\ee
through the field equation $\partial_j f_{ij}=J_i$.
The shift in the sphaleron energy due to the dipole moment interaction is given by~\cite{Manton}
\be
E_{\rm dipole}=-\int d^3 x \,\, a_i\, J_i\,,
\ee
which is negative. 
At the first order in $\theta_{\rm w}$ one can neglect $a_i$ in the current which then becomes
\be
J_i^{(1)}=-{1\over 2} {g' v^2}{h^2(gvr)(1-f(gvr))\over r^2}\epsilon_{3ij}x_j  \,,
\ee
where the sphaleron radial functions $f(\xi),h(\xi)$ ($\xi=g v r$) are the solutions of
 Eqs.~(\ref{fpp})-(\ref{hpp}) at $T\neq 0$.
Using the vector potential of a constant magnetic field $B$ along
the $\hat z$-axis,  $a_i=-(B/2) \epsilon_{3ij}x_j$, the dipole energy reads
\be
E_{\rm dipole}^{(1)}(T,B)=-\mu(T) \,B\,,
\label{Edipole}
\ee
where
\be
\mu(T)={2\pi\over 3}{g'\over g^3 v(T)} \int_0^\infty d\xi \xi^2 h^2(\xi)[1-f(\xi)]\,,
\label{mu}
\ee
and the temperature dependence also resides in the radial functions $f(\xi),h(\xi)$.

%%%%%%%%%%%%%%%%%%%%%%%%%%%%%%%%%%%%%
\bibliographystyle{JHEP}

\begin{thebibliography}{99}

\bibitem{baryogenesis} 
  A.~D.~Sakharov,
%  ``Violation of CP Invariance, c Asymmetry, and Baryon Asymmetry of the Universe,''
  Pisma Zh.\ Eksp.\ Teor.\ Fiz.\  {\bf 5}, 32-35 (1967).


\bibitem{reviews} For some reviews, see:
 A.~G.~Cohen, D.~B.~Kaplan and A.~E.~Nelson,
  %``Progress in electroweak baryogenesis,''
  Ann.\ Rev.\ Nucl.\ Part.\ Sci.\  {\bf 43}, 27 (1993)
                                     \href{http://arXiv.org/abs/hep-ph/9302210}{[hep-ph/9302210]};
  %%CITATION = ARNUA,43,27;%%
  M.~Quiros,
  %``Field theory at finite temperature and phase transitions,''
  Helv.\ Phys.\ Acta {\bf 67}, 451 (1994);
  %%CITATION = APPOA,B38,3661;%%
  V.~A.~Rubakov and M.~E.~Shaposhnikov,
  %``Electroweak baryon number non-conservation in the early universe and in
  %high-energy collisions,''
  Usp.\ Fiz.\ Nauk {\bf 166}, 493 (1996)
  [Phys.\ Usp.\  {\bf 39}, 461 (1996)]
                                     \href{http://arXiv.org/abs/hep-ph/9603208}{[hep-ph/9603208]};
  %%CITATION = PHUSE,39,461;%%
  M.~S.~Carena and C.~E.~M.~Wagner
  %``Electroweak baryogenesis and Higgs physics,''
                                      \href{http://arXiv.org/abs/hep-ph/9704347}{[hep-ph/9704347]};
  %%CITATION = HEP-PH/9704347;%%
  M.~Quiros
  %``Finite temperature field theory and phase transitions,''
  \href{http://arXiv.org/abs/hep-ph/9901312}{[hep-ph/9901312]}.
  %%CITATION = HEP-PH/9901312;%%


\bibitem{sphalerons} 
P.~Arnold and L.~D.~McLerran,
  %``Sphalerons, Small Fluctuations And Baryon Number Violation In Electroweak
  %Theory,''
  Phys.\ Rev.\  D {\bf 36}, 581 (1987);
  %%CITATION = PHRVA,D36,581;%%
%``The Sphaleron Strikes Back,''
  Phys.\ Rev.\  D {\bf 37}, 1020 (1988);
  %%CITATION = PHRVA,D37,1020;%%
  S.~Y.~Khlebnikov and M.~E.~Shaposhnikov,
  %``The Statistical Theory of Anomalous Fermion Number Nonconservation,''
  Nucl.\ Phys.\  B {\bf 308}, 885 (1988);
  %%CITATION = NUPHA,B308,885;%%
  F.~R.~Klinkhamer and N.~S.~Manton,
  %``A Saddle Point Solution In The Weinberg-Salam Theory,''
  Phys.\ Rev.\  D {\bf 30}, 2212 (1984);
  %%CITATION = PHRVA,D30,2212;%%
  B.~M.~Kastening, R.~D.~Peccei and X.~Zhang,
  %``Sphalerons in the two doublet Higgs model,''
  Phys.\ Lett.\  B {\bf 266}, 413 (1991);
  %%CITATION = PHLTA,B266,413;%%
  L.~Carson, X.~Li, L.~D.~McLerran and R.~T.~Wang,
  %``EXACT COMPUTATION OF THE SMALL FLUCTUATION DETERMINANT AROUND A
  %SPHALERON,''
  Phys.\ Rev.\  D {\bf 42}, 2127 (1990);
  %%CITATION = PHRVA,D42,2127;%%
  M.~Dine, P.~Huet and R.~L.~Singleton,
  %``Baryogenesis at the electroweak scale,''
  Nucl.\ Phys.\  B {\bf 375}, 625 (1992).
  %%CITATION = NUPHA,B375,625;%%

  
\bibitem{ereview}
  For a review, see
   K.~Enqvist,
  %``Primordial magnetic fields,''
   Int.\ J.\ Mod.\ Phys.\ D {\bf 7}, 331 (1998)
   \href{http://arXiv.org/abs/astro-ph/9803196}{[astro-ph/9803196]}.


\bibitem{Stevens}
  T.~Stevens and M.~B.~Johnson,
  %``Theory of Magnetic Seed-Field Theory of Magnetic Seed-Field Generation
  %during the Cosmological First-Order Electroweak Phase Transition,''
  %%CITATION = ARXIV:1001.3694;%%
 \href{http://arXiv.org/abs/arXiv:1001.3694}{[arXiv:1001.3694]};
  
\bibitem{baym} G.~Baym, D.~Bodeker and L.~D.~McLerran,
  %``Magnetic fields produced by phase transition bubbles in the electroweak
  %phase transition,''
  Phys.\ Rev.\ D {\bf 53}, 662 (1996)
  \href{http://arXiv.org/abs/hep-ph/9507429}{[hep-ph/9507429]}.
                                                                            

\bibitem{sigl} G.~Sigl, A.~V.~Olinto and K.~Jedamzik,
  %``Primordial magnetic fields from cosmological first order phase
  %transitions,''
  Phys.\ Rev.\ D {\bf 55}, 4582 (1997)
  \href{http://arXiv.org/abs/astro-ph/9610201}{[astro-ph/9610201]}.
                                                                                                                                                    
                                                                          
\bibitem{kibble} T. W. B. Kibble and A. Vilenkin, Phys. Rev. D {\bf
    52}, 679 (1995)
  \href{http://arXiv.org/abs/hep-ph/9501266}{[hep-ph/9501266]}.
                                                                               
  
 \bibitem{ae}
  J.~Ahonen and K.~Enqvist,
  %``Magnetic field generation in first order phase transition bubble
  %collisions,''
  Phys.\ Rev.\ D {\bf 57}, 664 (1998)
  \href{http://arXiv.org/abs/hep-ph/9704334}{[hep-ph/9704334]}.
                                                                              
                                                                              

\bibitem{con} J. Ahonen and K. Enqvist, Phys. Lett. B {\bf 382}, 40 (1996).

\bibitem{caprini}
  C.~Caprini,
  %``Limits for primordial magnetic fields,''
  PoS {\bf TEXAS2010}, 222 (2010)
  \href{http://arxiv.org/abs/1103.4060}{[arXiv:1103.4060]}.


\bibitem{Manton}
  F.~R.~Klinkhamer, N.~S. Manton,
  Phys.\ Rev. D  {\bf 30}, 2212 (1984).    
  
\bibitem{CGPR}
  D.~Comelli, D.~Grasso, M.~Pietroni, A.~Riotto,
  Phys.\ Lett. B\  {\bf 458}, 304 (1999)
                                   \href{http://arXiv.org/abs/hep-ph/9903227}{[hep-ph/9903227]}.
  
  
\bibitem{AndH} 
G.~W.~Anderson and L.~J.~Hall,
  %``The Electroweak Phase Transition And Baryogenesis,''
  Phys.\ Rev.\  D {\bf 45}, 2685 (1992).
  %%CITATION = PHRVA,D45,2685;%%

\bibitem{improvement}
M.~E.~Carrington,
  %``The Effective potential at finite temperature in the Standard Model,''
  Phys.\ Rev.\  D {\bf 45}, 2933 (1992);
  %%CITATION = PHRVA,D45,2933;%%
  M.~Dine, R.~G.~Leigh, P.~Huet, A.~D.~Linde and D.~A.~Linde,
  %``Comments on the electroweak phase transition,''
  Phys.\ Lett.\  B {\bf 283}, 319 (1992)
                                       \href{http://arXiv.org/abs/hep-ph/9203201}{[hep-ph/9203201]};
  %%CITATION = PHLTA,B283,319;%%
  Phys.\ Rev.\  D {\bf 46}, 550 (1992)
                                         \href{http://arXiv.org/abs/hep-ph/9203203}{[hep-ph/9203203]};
  %%CITATION = PHRVA,D46,550;%% 
  J.~R.~Espinosa, M.~Quiros and F.~Zwirner,
  %``On the nature of the electroweak phase transition,''
  Phys.\ Lett.\  B {\bf 314}, 206 (1993)
                                           \href{http://arXiv.org/abs/hep-ph/9212248}{[hep-ph/9212248]};
  %%CITATION = PHLTA,B314,206;%%
  W.~Buchmuller, Z.~Fodor, T.~Helbig and D.~Walliser,
  %``The Weak - electroweak phase transition,''
  Annals Phys.\  {\bf 234}, 260 (1994)
                                             \href{http://arXiv.org/abs/hep-ph/9303251}{[hep-ph/9303251]}.
  %%CITATION = APNYA,234,260;%%

%
\bibitem{twoloop} 
P.~Arnold and O.~Espinosa,
  %``The Effective potential and first order phase transitions: Beyond
  %leading-order,''
  Phys.\ Rev.\  D {\bf 47}, 3546 (1993)
  [Erratum-ibid.\  D {\bf 50}, 6662 (1994)]
                                               \href{http://arXiv.org/abs/hep-ph/9212235}{[hep-ph/9212235]}.
  %%CITATION = PHRVA,D47,3546;%%
  
  
  \bibitem{nonpert}
K.~Kajantie, K.~Rummukainen and M.~E.~Shaposhnikov,
  %``A Lattice Monte Carlo Study Of The Hot Electroweak Phase Transition,''
  Nucl.\ Phys.\  B {\bf 407}, 356 (1993)
                                                \href{http://arXiv.org/abs/hep-ph/9305345}{[hep-ph/9305345]};
  %%CITATION = NUPHA,B407,356;%%
  Z.~Fodor, J.~Hein, K.~Jansen, A.~Jaster and I.~Montvay,
  %``Simulating the electroweak phase transition in the SU(2) Higgs model,''
  Nucl.\ Phys.\  B {\bf 439}, 147 (1995)
                                                  \href{http://arXiv.org/abs/hep-lat/9409017}{[hep-lat/9409017]};
  %%CITATION = NUPHA,B439,147;%%
  K.~Kajantie, M.~Laine, K.~Rummukainen and M.~E.~Shaposhnikov,
  %``The Electroweak Phase Transition: A Non-Perturbative Analysis,''
  Nucl.\ Phys.\  B {\bf 466}, 189 (1996)
                                                    \href{http://arXiv.org/abs/hep-lat/9510020}{[hep-lat/9510020]};
  %%CITATION = NUPHA,B466,189;%%
  K.~Jansen,
  %``Status of the Finite Temperature Electroweak Phase Transition on the
  %Lattice,''
  Nucl.\ Phys.\ Proc.\ Suppl.\  {\bf 47}, 196 (1996)
                                                      \href{http://arXiv.org/abs/hep-lat/9509018}{[hep-lat/9509018]}.
  %%CITATION = NUPHZ,47,196;%%
For an alternative approach, see:
 B.~Bergerhoff and C.~Wetterich,
  %``The Strongly Interacting Electroweak Phase Transition,''
  Nucl.\ Phys.\  B {\bf 440}, 171 (1995)
                                                      \href{http://arXiv.org/abs/hep-ph/9409295}{[hep-ph/9409295]}
 and references therein.
  %%CITATION = NUPHA,B440,171;%%

  \bibitem{CPSM}
 G.~R.~Farrar and M.~E.~Shaposhnikov,
  %``Baryon Asymmetry Of The Universe In The Minimal Standard Model,''
  Phys.\ Rev.\ Lett.\  {\bf 70}, 2833 (1993)
  [Erratum-ibid.\  {\bf 71}, 210 (1993)]
                                                        \href{http://arXiv.org/abs/hep-ph/9305274}{[hep-ph/9305274]};
  %%CITATION = PRLTA,70,2833;%% 
  M.~B.~Gavela, P.~Hernandez, J.~Orloff and O.~Pene,
  %``Standard model CP violation and baryon asymmetry,''
  Mod.\ Phys.\ Lett.\  A {\bf 9}, 795 (1994)
                                                          \href{http://arXiv.org/abs/hep-ph/9312215}{[hep-ph/9312215]};
  %%CITATION = MPLAE,A9,795;%%
  M.~B.~Gavela, P.~Hernandez, J.~Orloff, O.~Pene and C.~Quimbay,
  %``Standard model CP violation and baryon asymmetry. Part 2: Finite
  %temperature,''
  Nucl.\ Phys.\  B {\bf 430}, 382 (1994)
                                                            \href{http://arXiv.org/abs/hep-ph/9406289}{[hep-ph/9406289]};
  %%CITATION = NUPHA,B430,382;%%
  P.~Huet and E.~Sather,
  %``Electroweak baryogenesis and standard model CP violation,''
  Phys.\ Rev.\  D {\bf 51}, 379 (1995)
                                                              \href{http://arXiv.org/abs/hep-ph/9404302}{[hep-ph/9404302]}.
  %%CITATION = PHRVA,D51,379;%%

  
  
  \bibitem{early} 
 G.~F.~Giudice,
  %``The Electroweak phase transition in supersymmetry,''
  Phys.\ Rev.\  D {\bf 45}, 3177 (1992);
  %%CITATION = PHRVA,D45,3177;%% 
  K.~S.~Myint,
  %``Double Strangeness Five-Body System,''
  Nucl.\ Phys.\  A {\bf 547}, 227C (1992).
  %%CITATION = NUPHA,A547,227C;%%

%
\bibitem{mariano1} 
 J.~R.~Espinosa, M.~Quiros and F.~Zwirner,
  %``On the electroweak phase transition in the minimal supersymmetric Standard
  %Model,''
  Phys.\ Lett.\  B {\bf 307}, 106 (1993)
                                                                \href{http://arXiv.org/abs/hep-ph/9303317}{[hep-ph/9303317]}.
  %%CITATION = PHLTA,B307,106;%%
%
\bibitem{mariano2}
A.~Brignole, J.~R.~Espinosa, M.~Quiros and F.~Zwirner,
  %``Aspects Of The Electroweak Phase Transition In The Minimal Supersymmetric
  %Standard Model,''
  Phys.\ Lett.\  B {\bf 324}, 181 (1994)
                                                                  \href{http://arXiv.org/abs/hep-ph/9312296}{[hep-ph/9312296]}.
  %%CITATION = PHLTA,B324,181;%%


\bibitem{CQW} 
M.~S.~Carena, M.~Quiros and C.~E.~M.~Wagner,
  %``Opening the Window for Electroweak Baryogenesis,''
  Phys.\ Lett.\  B {\bf 380}, 81 (1996)
                                                                  \href{http://arXiv.org/abs/hep-ph/9603420}{[hep-ph/9603420]}. 
  %%CITATION = PHLTA,B380,81;%%

%
\bibitem{Delepine} 
D.~Delepine, J.~M.~Gerard, R.~Gonzalez Felipe and J.~Weyers,
  %``A light stop and electroweak baryogenesis,''
  Phys.\ Lett.\  B {\bf 386}, 183 (1996)
                                                                    \href{http://arXiv.org/abs/hep-ph/9604440}{[hep-ph/9604440]}. 
  %%CITATION = PHLTA,B386,183;%%


%
\bibitem{CK} 
 J.~M.~Cline and K.~Kainulainen,
  %``Supersymmetric Electroweak Phase Transition: Beyond Perturbation Theory,''
  Nucl.\ Phys.\  B {\bf 482}, 73 (1996)
                                                                      \href{http://arXiv.org/abs/hep-ph/9605235}{[hep-ph/9605235]};
  %%CITATION = NUPHA,B482,73;%%
  %``Supersymmetric electroweak phase transition: Dimensional reduction  versus
  %effective potential,''
  Nucl.\ Phys.\  B {\bf 510}, 88 (1998)
                                                                        \href{http://arXiv.org/abs/hep-ph/9705201}{[hep-ph/9705201]};
  %%CITATION = NUPHA,B510,88;%%
 M.~Laine and K.~Rummukainen,
  %``The MSSM electroweak phase transition on the lattice,''
  Nucl.\ Phys.\  B {\bf 535}, 423 (1998)
                                                                        \href{http://arXiv.org/abs/hep-lat/9804019}{[hep-lat/9804019]};
  %%CITATION = NUPHA,B535,423;%%
%M.~Laine and K.~Rummukainen,
  %``A strong electroweak phase transition up to m(H) approx. 105-GeV,''
  Phys.\ Rev.\ Lett.\  {\bf 80}, 5259 (1998)
                                                                          \href{http://arXiv.org/abs/hep-ph/9804255}{[hep-ph/9804255]}.
  %%CITATION = PRLTA,80,5259;%%
%

\bibitem{FL} 
 M.~Laine,
  %``Effective theories of MSSM at high temperature,''
  Nucl.\ Phys.\  B {\bf 481}, 43 (1996)
  [Erratum-ibid.\  B {\bf 548}, 637 (1999)]
                                                                            \href{http://arXiv.org/abs/hep-ph/9605283}{[hep-ph/9605283]};
  %%CITATION = NUPHA,B481,43;%%
M.~Losada,
  %``High temperature dimensional reduction of the MSSM and other  multi-scalar
  %models,''
  Phys.\ Rev.\  D {\bf 56}, 2893 (1997)
                                                                             \href{http://arXiv.org/abs/hep-ph/9605266}{[hep-ph/9605266]};
  %%CITATION = PHRVA,D56,2893;%%
  %``The electroweak phase transition in the minimal supersymmetric standard
  %model,''
%  preprint arXiv:hep-ph/9612337;
  %%CITATION = HEP-PH/9612337;%%
G.~R.~Farrar and M.~Losada,
  %``SUSY and the electroweak phase transition,''
  Phys.\ Lett.\  B {\bf 406}, 60 (1997)
                                                                               \href{http://arXiv.org/abs/hep-ph/9612346}{[hep-ph/9612346]}.
  %%CITATION = PHLTA,B406,60;%%
%
\bibitem{JoseR}
J.~R.~Espinosa,
  %``Dominant Two-Loop Corrections to the MSSM Finite Temperature Effective
  %Potential,''
  Nucl.\ Phys.\  B {\bf 475}, 273 (1996)
                                                                                 \href{http://arXiv.org/abs/hep-ph/9604320}{[hep-ph/9604320]}.
  %%CITATION = NUPHA,B475,273;%%
%
\bibitem{JRB} 
 B.~de Carlos and J.~R.~Espinosa,
  %``The baryogenesis window in the MSSM,''
  Nucl.\ Phys.\  B {\bf 503}, 24 (1997)
                                                                                 \href{http://arXiv.org/abs/hep-ph/9703212}{[hep-ph/9703212]}.  
  %%CITATION = NUPHA,B503,24;%%
%
\bibitem{Carena:1997gx}
M.~S.~Carena, M.~Quiros, A.~Riotto, I.~Vilja and C.~E.~M.~Wagner,
  %``Electroweak baryogenesis and low energy supersymmetry,''
  Nucl.\ Phys.\  B {\bf 503}, 387 (1997)
                                                                                  \href{http://arXiv.org/abs/hep-ph/9702409}{[hep-ph/9702409]}.  
  %%CITATION = NUPHA,B503,387;%%

%
\bibitem{Carena:1997ki}
  M.~S.~Carena, M.~Quiros and C.~E.~M.~Wagner,
  %``Electroweak baryogenesis and Higgs and stop searches at LEP and the
  %Tevatron,''
  Nucl.\ Phys.\  B {\bf 524}, 3 (1998)
                                                                                 \href{http://arXiv.org/abs/hep-ph/9710401}{[hep-ph/9710401]}.  
  %%CITATION = NUPHA,B524,3;%%
%
\bibitem{CJK}
 J.~M.~Cline, M.~Joyce and K.~Kainulainen,
  %``Supersymmetric electroweak baryogenesis in the WKB approximation,''
  Phys.\ Lett.\  B {\bf 417}, 79 (1998)
  [Erratum-ibid.\  B {\bf 448}, 321 (1999)]
                                                                                 \href{http://arXiv.org/abs/hep-ph/9708393}{[hep-ph/9708393]}.  
  %%CITATION = PHLTA,B417,79;%%
%
\bibitem{Iiro2} 
T.~Multamaki and I.~Vilja,
  %``CP-violation and baryogenesis in the low energy minimal supersymmetric
  %standard model,''
  Phys.\ Lett.\  B {\bf 411}, 301 (1997)
                                                                                 \href{http://arXiv.org/abs/hep-ph/9705469}{[hep-ph/9705469]}.    
%
\bibitem{Toni2} 
A.~Riotto,
  %``More about electroweak baryogenesis in the minimal supersymmetric  standard
  %model,''
  Int.\ J.\ Mod.\ Phys.\  D {\bf 7}, 815 (1998)
                                                                                  \href{http://arXiv.org/abs/hep-ph/9709286}{[hep-ph/9709286]}.    
  %%CITATION = IMPAE,D7,815;%%
%

\bibitem{Toni3} 
A.~Riotto,
  %``Supersymmetric electroweak baryogenesis, nonequilibrium field theory and
  %quantum Boltzmann equations,''
  Nucl.\ Phys.\  B {\bf 518}, 339 (1998)
                                                                                    \href{http://arXiv.org/abs/hep-ph/9712221}{[hep-ph/9712221]}.   




\bibitem{Worah} 
 M.~P.~Worah,
  %``Supersymmetric baryogenesis at the electroweak phase transition,''
  Phys.\ Rev.\  D {\bf 56}, 2010 (1997)
                                                                                  \href{http://arXiv.org/abs/hep-ph/9702423}{[hep-ph/9702423]}.      
  %%CITATION = PHRVA,D56,2010;%%
%
\bibitem{Schmidt}
  D.~Bodeker, P.~John, M.~Laine and M.~G.~Schmidt,
  %``The 2-loop MSSM finite temperature effective potential with stop
  %condensation,''
  Nucl.\ Phys.\  B {\bf 497}, 387 (1997)
                                                                          \href{http://arXiv.org/abs/hep-ph/9612364}{[hep-ph/9612364]}.
  %%CITATION = NUPHA,B497,387;%%
%

\bibitem{Cline:2000kb}
  J.~M.~Cline and K.~Kainulainen,
  %``A new source for electroweak baryogenesis in the MSSM,''
  Phys.\ Rev.\ Lett.\  {\bf 85}, 5519 (2000)
                                                                          \href{http://arXiv.org/abs/hep-ph/0002272}{[hep-ph/0002272]}; 
  %%CITATION = PRLTA,85,5519;%%
 J.~M.~Cline, M.~Joyce and K.~Kainulainen,
  %``Supersymmetric electroweak baryogenesis,''
  JHEP {\bf 0007}, 018 (2000)
                                                                            \href{http://arXiv.org/abs/hep-ph/0006119}{[hep-ph/0006119]}.  
  %%CITATION = JHEPA,0007,018;%%

\bibitem{Carena:2000id}
  M.~S.~Carena, J.~M.~Moreno, M.~Quiros, M.~Seco and C.~E.~M.~Wagner,
  %``Supersymmetric CP-violating currents and electroweak baryogenesis,''
  Nucl.\ Phys.\  B {\bf 599}, 158 (2001)
                                                                             \href{http://arXiv.org/abs/hep-ph/0011055}{[hep-ph/0011055]};
  %%CITATION = NUPHA,B599,158;%%
 M.~S.~Carena, M.~Quiros, M.~Seco and C.~E.~M.~Wagner,
  %``Improved results in supersymmetric electroweak baryogenesis,''
  Nucl.\ Phys.\  B {\bf 650}, 24 (2003)
                                                                               \href{http://arXiv.org/abs/hep-ph/0208043}{[hep-ph/0208043]}.
  %%CITATION = NUPHA,B650,24;%%

\bibitem{Konstandin:2005cd}
  T.~Konstandin, T.~Prokopec, M.~G.~Schmidt and M.~Seco,
  %``MSSM electroweak baryogenesis and flavour mixing in transport  equations,''
  Nucl.\ Phys.\ B {\bf 738}, 1 (2006)
                                                                               \href{http://arXiv.org/abs/hep-ph/0505103}{[hep-ph/0505103]}.  
  %%CITATION = HEP-PH 0505103;%%

\bibitem{Cirigliano:2006dg}
C.~Lee, V.~Cirigliano and M.~J.~Ramsey-Musolf,
  %``Resonant relaxation in electroweak baryogenesis,''
  Phys.\ Rev.\  D {\bf 71}, 075010 (2005)
 \href{http://arXiv.org/abs/hep-ph/0412354}{[hep-ph/0412354]};
  %%CITATION = PHRVA,D71,075010;%%
  V.~Cirigliano, S.~Profumo and M.~J.~Ramsey-Musolf,
  %``Baryogenesis, electric dipole moments and dark matter in the MSSM,''
  JHEP {\bf 0607}, 002 (2006)
                                                                                 \href{http://arXiv.org/abs/hep-ph/0603246}{[hep-ph/0603246]};  
  %%CITATION = JHEPA,0607,002;%%
V.~Cirigliano, C.~Lee and S.~Tulin
  %``Resonant Flavor Oscillations in Electroweak Baryogenesis,''
    %%CITATION = ARXIV:1106.0747;%%
 \href{http://arXiv.org/abs/arXiv:1106.0747}{[arXiv:1106.0747]}.  

\bibitem{qn} M.~Carena, G.~Nardini, M.~Quiros and C.~E.~M.~Wagner,
  %``The Effective Theory of the Light Stop Scenario,''
  JHEP {\bf 0810}, 062 (2008)
                                                                                   \href{http://arXiv.org/abs/0806.4297}{[arXiv:0806.4297]}.  

  


\bibitem{chung1} D.~J.~H.~Chung, B.~Garbrecht and S.~Tulin,
  %``The Effect of the Sparticle Mass Spectrum on the Conversion of B-L to B,''
  JCAP {\bf 0903}, 008 (2009)
                                                                                     \href{http://arXiv.org/abs/0807.2283}{[arXiv:0807.2283]}.  

\bibitem{chung2} D.~J.~H.~Chung, B.~Garbrecht, M.~J.~Ramsey-Musolf and S.~Tulin,
  %``Yukawa Interactions and Supersymmetric Electroweak Baryogenesis,''
  Phys.\ Rev.\ Lett.\  {\bf 102}, 061301 (2009)
                                                                                       \href{http://arXiv.org/abs/0808.1144}{[arXiv:0808.1144]}.  


\bibitem{chung3} D.~J.~H.~Chung, B.~Garbrecht, M.~J.~Ramsey-Musolf and S.~Tulin,
  %``Supergauge interactions and electroweak baryogenesis,''
  JHEP {\bf 0912}, 067 (2009)
                                                                                         \href{http://arXiv.org/abs/0908.2187}{[arXiv:0908.2187]}.  
 
 
\bibitem{window}
  M.~Carena, G.~Nardini, M.~Quiros, C.~E.~M.~Wagner,
  Nucl.\ Phys.\  B{\bf 812}, 243, (2009)
                                   \href{http://arXiv.org/abs/0809.3760}{[arXiv:0809.3760]}.


\bibitem{EDM} 
A.~Pilaftsis,
  %``Higgs-mediated electric dipole moments in the MSSM: An application to
  %baryogenesis and Higgs searches,''
  Nucl.\ Phys.\  B {\bf 644}, 263 (2002)
    \href{http://arXiv.org/abs/hep-ph/0207277}{[hep-ph/0207277]};
  %%CITATION = NUPHA,B644,263;%%
D.~Chang, W.~F.~Chang and W.~Y.~Keung,
  %``Electric dipole moment in the split supersymmetry models,''
  Phys.\ Rev.\  D {\bf 71}, 076006 (2005)
    \href{http://arXiv.org/abs/hep-ph/0503055}{[hep-ph/0503055]};
  %%CITATION = PHRVA,D71,076006;%%
G.~F.~Giudice and A.~Romanino,
  %``Electric dipole moments in split supersymmetry,''
  Phys.\ Lett.\  B {\bf 634}, 307 (2006)
    \href{http://arXiv.org/abs/hep-ph/0510197}{[hep-ph/0510197]};
  %%CITATION = PHLTA,B634,307;%%
 Y.~Li, S.~Profumo and M.~Ramsey-Musolf,
  %``Higgs-Higgsino-Gaugino Induced Two Loop Electric Dipole Moments,''
  Phys.\ Rev.\  D {\bf 78}, 075009 (2008)
  \href{http://arXiv.org/abs/0806.2693}{[arXiv:0806.2693]};
  %%CITATION = PHRVA,D78,075009;%%
%
   V.~Cirigliano, Y.~Li, S.~Profumo and M.~J.~Ramsey-Musolf,
  %``MSSM Baryogenesis and Electric Dipole Moments: An Update on the
  %Phenomenology,''
  JHEP {\bf 1001}, 002 (2010)
 \href{http://arXiv.org/abs/0910.4589}{[arXiv:0910.4589]};
  %%CITATION = JHEPA,1001,002;%%
%
  Y.~Li, S.~Profumo and M.~Ramsey-Musolf,
  %``A Comprehensive Analysis of Electric Dipole Moment Constraints on
  %CP-violating Phases in the MSSM,''
  JHEP {\bf 1008}, 062 (2010)
 \href{http://arXiv.org/abs/1006.1440}{[arXiv:1006.1440]}.
  %%CITATION = JHEPA,1008,062;%%

%\bibitem{carena} {\bf To FILL IN}.

\bibitem{lep} 
R.~Barate {\it et al.} [LEP Working Group for Higgs boson searches and ALEPH and DELPHI and L3 and OPAL Collaborations],
  %``Search for the standard model Higgs boson at LEP,''
  Phys.\ Lett.\  {\bf B565}, 61-75 (2003)
                                   \href{http://arXiv.org/abs/hep-ex/0306033}{[hep-ex/0306033]}.

\bibitem{Linde} 
  A.~D.~Linde,
  %``Decay Of The False Vacuum At Finite Temperature,''
  Nucl.\ Phys.\  B {\bf 216} 421 (1983)
  [Erratum-ibid.\  B {\bf 223} 544 (1983)].
  %%CITATION = NUPHA,B216,421;%%

\bibitem{thomas}
  A.~H.~Guth and S.~H.~H.~Tye,
  %``Phase Transitions And Magnetic Monopole Production In The Very Early
  %Universe,''
  Phys.\ Rev.\ Lett.\  {\bf 44}, 631 (1980)
  [Erratum-ibid.\  {\bf 44}, 963 (1980)],
  %%CITATION = PRLTA,44,631;%%
  A.~H.~Guth and E.~J.~Weinberg,
  %``Cosmological Consequences Of A First Order Phase Transition In The SU(5)
  %Grand Unified Model,''
  Phys.\ Rev.\  D {\bf 23}, 876 (1981),
  %%CITATION = PHRVA,D23,876;%%
  S.~J.~Huber and T.~Konstandin,
  %``Production of Gravitational Waves in the nMSSM,''
  JCAP {\bf 0805}, 017 (2008)
   \href{http://arXiv.org/abs/0709.2091}{[arXiv:0709.2091]}.
  %%CITATION = JCAPA,0805,017;%%


\bibitem{cdf}
   T.Aaltonen et al., the CDF Collaboration, CDF Public Note 9834 (2009).

\bibitem{Navarro}
  P.~Elmfors, K.~Enqvist, K.~Kainulainen,
  %``Strongly first order electroweak phase transition induced by primordial hypermagnetic fields,''
  Phys.\ Lett.\  {\bf B440}, 269-274 (1998)
     \href{http://arXiv.org/abs/hep-ph/9806403}{[hep-ph/9806403]};
     J.~Navarro, A.~Sanchez, M.~E.~Tejeda-Yeomans, A.~Ayala and G.~Piccinelli,
  %``Symmetry restoration at finite temperature with weak magnetic fields,''
  Phys.\ Rev.\  D {\bf 82} (2010) 123007
\href{http://arXiv.org/abs/1007.4208}{[arXiv:1007.4208]}.
  %%CITATION = PHRVA,D82,123007;%%

\bibitem{gaugedep}
  H.~H.~Patel, M.~J.~Ramsey-Musolf,
  %``Baryon Washout, Electroweak Phase Transition, and Perturbation Theory,''
  JHEP {\bf 1107}, 029 (2011)
\href{http://arXiv.org/abs/1101.4665}{[arXiv:1101.4665]};
 C.~Wainwright, S.~Profumo, M.~J.~Ramsey-Musolf,
  %``Gravity Waves from a Cosmological Phase Transition: Gauge Artifacts and Daisy Resummations,''
  Phys.\ Rev.\  {\bf D84}, 023521 (2011)
  \href{http://arXiv.org/abs/1104.5487}{[arXiv:1104.5487]}.
  



\bibitem{Laterveer}
  F.~R.~Klinkhamer, R.~Laterveer,
  Z.\ Phys.\  C{\bf 53}, 247(1992).
  

\bibitem{sphaleronMSSM}
  J.~M.~Moreno, D.~H.~Oaknin and M.~Quiros,
  Nucl.\ Phys.\  B {\bf 483}, 267 (1997)
                                   \href{http://arXiv.org/abs/hep-ph//9605387}{[hep-ph/9605387]}.
                                   
                                   
\bibitem{sphaleron_rate}
  M.~E.~Shaposhnikov,
  JETP Lett.\  {\bf 44}, 465 (1986);
  M.~E.~Shaposhnikov,
  Nucl.\ Phys.\  B{\bf 287}, 757 (1987).
  

\bibitem{DHN}
  R.~F.~Dashen, B.~Hasslacher, A.~Neveu,
  Phys.\ Rev.\  D{\bf 10}, 4138 (1974);
  P.~Forgacs, Z.~Horvath,
  Phys.\ Lett.\  B{\bf 138}, 397 (1984).
\bibitem{seven_functions}
  B.~Kleihaus, J.~Kunz, Y.~Brihaye,
  Phys.\ Lett.\  B{\bf 273}, 100 (1991);
    J.~Kunz, B.~Kleihaus, Y.~Brihaye,
  Phys.\ Rev.\ D  {\bf 46}, 3587 (1992).
   

  
                                 


\end{thebibliography}

%%%%%%%%%%%%%%%%%%%%%%%%%%%%%%%%%%%%%

\end{document}